\documentclass[12pt,preprint]{aastex}

\begin{document}

\title{Rings and Bent Chain Galaxies in the GEMS and GOOD Fields}

\author{Debra Meloy Elmegreen \affil{Vassar College, Dept. of
Physics \& Astronomy, Box 745, Poughkeepsie, NY 12604;
elmegreen@vassar.edu} and}
\author{Bruce G. Elmegreen\affil{IBM Research Division, T.J. Watson
Research Center, P.O. Box 218, Yorktown Heights, NY 10598,
bge@watson.ibm.com} }
\begin{abstract}
Twenty-four galaxies with rings or partial rings were studied in
the GEMS and GOODS fields out to z$\sim$1.4. Most resemble local
collisional ring galaxies in morphology, size, and clumpy star
formation. Clump ages range from $10^8$ to $10^9$ yr and clump
masses go up to several $\times10^8$ M$_\odot$, based on color
evolution models. The clump ages are consistent with the expected
lifetimes of ring structures if they are formed by collisions.
Fifteen other galaxies that resemble the arcs in partial ring
galaxies but have no evident disk emission were also studied.
Their clumps have bluer colors at all redshifts compared to the
clumps in the ring and partial ring sample, and their clump ages
are younger than in rings and partial rings by a factor of
$\sim10$. In most respects, they resemble chain galaxies except
for their curvature; we refer to them as ``bent chains.'' Several
rings are symmetric with centered nuclei and no obvious
companions. They could be outer Lindblad resonance rings, although
some have no obvious bars or spirals to drive them. If these
symmetric cases are resonance rings, then they could be the
precursors of modern resonance rings, which are only $\sim30$\%
larger on average. This similarity in radius suggests that the
driving pattern speed has not slowed by more by $\sim30$\% during
the last $\sim7$ Gy. Those without bars could be examples of
dissolved bars.
\end{abstract}


\section{Introduction}

Outer rings in galaxies can be the result of plunging impacts,
spiral and bar resonances, or satellite accretion onto a stable
polar orbit. All types combined comprise less than 0.2\% of local
spiral galaxies (Athanassoula \& Bosma 1985), although 4\% of
early type galaxies in de Vaucouleurs \& Buta (1980) have outer
resonance rings. Galaxies with centrally located nuclei and smooth
outer rings are sometimes referred to as ``O-type,'' while those
with knotty structure and occasionally offset nuclei are called
``P-type'' (Few \& Madore 1986).

Collisional ring galaxies usually have a clumpy ring of star
formation that is disk material compressed by a wave driven
outward in response to the impacting galaxy's gravitational force
(Lynds \& Toomre 1976; Theys \& Spiegel 1976, 1977; Appleton \&
Struck-Marcell 1987; Struck 1997). Generally the intruder is
visible nearby and has a mass of at least $\sim10$\% of the main
galaxy mass (Lavery et al. 2004). Collisional ring galaxies may
have nuclei or offset nuclei, or they could be empty inside with
one prominent clump on the ring. All of these structures are
expected to last for only a brief time, $\sim10^8$ yrs (Theys \&
Spiegel 1976, 1977).

The most famous example of a collisional ring galaxy is the
Cartwheel (Fosbury \& Hawarden 1977; Struck et al. 1996), which
has an offset nucleus and a ring of recent star formation. A
catalog of southern galaxies by Arp \& Madore (1987) includes many
examples of collisional ring galaxies. Individual collisional ring
galaxies studied recently include Arp 107 (Smith et al. 2005), VII
Zw 466 (Appleton, Charmandaris, \& Struck 1996), AM0644-741
(Higdon \& Wallin 1997, NGC 922 (Wong et al. 2006), and a sample
of 11 galaxies including the Cartwheel (Appleton \& Marston 1997).
Numerical simulations of mergers have been used to reproduce many
features of these nearby systems (e.g., review by Barnes \&
Hernquist 1992; Hernquist \& Weil 1993; Mihos \& Hernquist 1994;
review by Struck 1999).

Resonance rings can be located near the nucleus, mid-disk, or
outer disk, depending on the resonance (see reviews by
Athanassoula \& Bosma 1985; Buta \& Combes 1996).  Among local
galaxies, the outer resonance rings prefer early-type spirals (de
Vaucouleurs \& Buta 1980). Polar rings are perpendicular to the
central galaxy's plane and usually associated with S0 types.

Collisional ring galaxies should be relatively more common in the
high redshift Universe because interactions were more frequent in
the past (e.g. Abraham et al. 1996a). The merger rate as a
function of redshift has been studied from visible morphologies
(Straughn et al. 2006), comoving densities of collisional ring
galaxies (Lavery et al. 2004), Gini coefficients (Lotz et al.
2006), angular correlation functions (Neuschaefer et al. 1997),
and asymmetry indices (Abraham et al. 1996b; Conselice et al.
2003). These studies find that the merger rate increases with
redshift as $(1+z)^n$ for $n$ between 2 and 7.

The GEMS (Galaxy Evolution from Morphology and SEDs; Rix et al.
2004), GOODS (Great Observatories Origins Deep Survey; Giavalisco
et al. 2004), and UDF (Ultra Deep Field; Beckwith et al. 2004)
surveys done with the HST ACS (Hubble Space Telescope Advanced
Camera for Surveys) have enabled unprecedented high resolution
studies of the morphology of intermediate and high redshift
galaxies. Here we examine the GEMS and GOODS fields for ring and
partial ring galaxies, and we compare the properties of these
galaxies to their local counterparts and to similar objects in the
UDF.

\section{Data and Analysis}

The UDF, GOODS, and GEMS images from the public archive were used
for this study. They include exposures in 4 filters for UDF and
GOODS:  F435W (B$_{435}$), F606W (V$_{606}$), F775W (i$_{775}$),
and F850LP (z$_{850}$); and 2 filters for GEMS (V and z). The
images were drizzled to produce final archival images with a scale
of 0.03 arcsec per px. GEMS, which incorporates the southern GOODS
survey (Chandra Deep Field South, CDF-S)  in the central quarter
of its field, covers 28 arcmin x 28 arcmin; there are 63 GEMS and
18 GOODS images that make up the whole field.  The GOODS images
have a limiting AB mag of V$_{606}$= 27.5 for an extended object,
or about two mags fainter than the GEMS images. The UDF is
equivalent to one of these subfields, but probes to a depth of
about 1.5 mag fainter than GOODS, or about 29 AB mag. There are
over 25,000 galaxies catalogued for GEMS in the Combo-17 survey
(Classifying Objects by Medium-Band Observations, a
spectrophotometric 17-filter survey; Wolf et al. 2003), as well as
over 10,000 galaxies in the UDF (Beckwith et al. 2004, on the
website archive.stsci.edu).

We examined the GEMS and GOODS fields for ring galaxies, first
using the online Skywalker images, and then using the 81 high
resolution V$_{606}$ fits images. We chose a limiting diameter of
$\sim20$ pixels because it is difficult to discern a ring
structure in smaller galaxies; in fact, all of our rings are
larger than 30 pixels. Snapshots of the ring galaxies in the
V$_{606}$ images are shown in Figure \ref{fig:bullsrings}, and
partial rings are in Figure \ref{fig:bullspartial}. Redshifts and
1 arcsec lengths are indicated.

The average redshift of our sample is 0.8$\pm$0.3, where a
V$_{606}$ image corresponds to the restframe U-band. We examined
all the available passbands for comparison and found little
difference in structure from band to band. The V$_{606}$ images in
the figures have the same rings, clumps, and nuclear structures as
the z$_{850}$ images, which correspond to the restframe B-band on
average. We show only the V$_{850}$ images because they have
higher signal-to-noise ratios. The bottom three galaxies in Figure
\ref{fig:bullspartial} have slightly different morphologies,
called RK (``ring-knot'') by Theys \& Spiegel (1976); they have a
very large clump on the ring (which may be the impacting galaxy)
and no large clump interior to the ring. In all, we selected 9
ring galaxies and 15 partial ring galaxies.

GEMS and GOODS redshifts were obtained from the COMBO-17 list
(Wolf et al. 2003). The data are in Table \ref{tab:rings}, which
gives the COMBO-17 catalog number and redshift for the galaxies in
order of presentation in the figures.  The redshift ranges from
$z=0.1$ to 1.4 in the surveyed area of $2.8\times10^6$
arcsec$^{2}$. For comparison, 25 collisional ring galaxies were
identified by Lavery et al. (2004) in 162 WFPC2 (Wide Field and
Planetary Camera) fields from z=0.1 to 1, covering a comparable
area of $4.2\times10^6$ arcsec$^{2}$. Our number per unit redshift
and per unit solid angle is similar to theirs.

In the GEMS and GOODS fields, 15 other galaxies were found to
resemble the clumpy arcs of partial ring galaxies, but they did
not have nuclei or inner disks suggestive of a ring galaxy. They
looked more like chain galaxies, which are common at intermediate
to high redshift (Cowie, Hu, \& Songaila 1995).  We refer to these
15 as ``bent chain'' galaxies. Their properties are listed in
Table \ref{tab:rings}, and they are shown in Figure
\ref{fig:bentchain}, with 2-$\sigma$ contour images in Figure
\ref{fig:contour2} that highlight the faintest structures. Some
bent chain galaxies could have suffered collisions like those in
collisional ring galaxies, as indicated by the wide variety of
shapes for direct hits shown by Struck (1999). In particular,
Struck simulates ``banana'' shapes reminiscent of bent chains, in
addition to partial rings reminiscent of the galaxies in Figure
\ref{fig:bullspartial}. It is not clear from the morphology alone
whether bent chains are the same as partial rings (we consider
this difference in more detail below).

To help understand these bent chains, we compared them to straight
chain galaxies and clump cluster galaxies, which are presumably
face-on versions of chain galaxies. We identified 19 chain
galaxies in the GEMS and GOODS field and added 29 chain and 44
clump cluster galaxies in the UDF. The UDF galaxies came from our
morphology catalog of the UDF, where we classified 884 galaxies
larger than 10 pixels in diameter into one of six main
morphological types: chain, double, tadpole, clump cluster,
spiral, or elliptical (Elmegreen et al. 2005a). The photometric
redshifts of the UDF galaxies were determined using the Bayesian
Photometric Redshift (BPZ) method (Benitez 2000; Coe et al. 2006;
Elmegreen et al. 2006). For the present study, we used only the
UDF chains and clump clusters with $z < 1$, because the galaxies
in our GEMS and GOODS surveys are primarily restricted to this
regime.

Ellipse fits were done on the ring galaxies using the IRAF task
{\it ellipse}. From these, average radial profiles were made to
assess the light and color distributions. The IRAF task {\it
pvector} was used to make intensity scans along the bent chain and
chain galaxies.

Photometry was done on each prominent clump or nucleus in the GEMS
and GOODS galaxies using the IRAF task {\it imstat}, in which a
box was defined around each clump to measure the number of pixels
and mean pixel count. Conversions to magnitudes were done using
the zeropoints tabulated in the ACS data handbook,
www.stsci.edu/hst/acs/analysis/zeropoints. Sky subtraction was not
done because the background was negligible. The photometric errors
are $\sim$0.1-0.2 mag for individual clumps.

\section{Results}
\subsection{Galaxy properties}

The COMBO-17 table included restframe U, B, and V magnitudes for
galaxies whose redshifts could be determined. Figure
\ref{fig:ringsf5} shows the absolute restframe V magnitude as a
function of redshift for the integrated galaxy light from rings
and partial rings (dots), bent chains (triangles), and chains
(circles) in the GEMS and GOODS fields (similar information has
not been tabulated for the UDF galaxies). The galaxies range from
absolute rest $V\sim-17$ to $-22$ mag. Three bent chains and one
partial ring galaxy were eliminated from this figure because their
small tabulated redshifts ($z<0.1$) gave them absolute magnitudes
of $-12$ to $-14$, which is not sensible considering their
similarity in appearance and angular size to the other galaxies in
our sample. The figure indicates that the ring and partial ring
galaxies are $\sim1.5$ mag brighter at each redshift than the bent
chain and chain galaxies.

The linear diameters of the galaxies (Table \ref{tab:rings}) were
determined from their redshifts and observed angular diameters
using the conversion for a $\Lambda$CDM cosmology (Carroll, Press,
\& Turner 1992; Spergel et al. 2003). They are shown on the left
in Figure \ref{fig:ringf10a} for the GEMS and GOODS galaxies (the
right-hand side of the figure will be discussed in the next
section). There is no correlation between minimum diameter and
redshift because the chosen galaxies were generally much larger
than our size cutoff. The drop in size for very small redshifts
could be the result of redshift errors, as discussed above, or it
could be the result of a selection of GOODS and GEMS fields that
avoid galaxies large in angular size. The range in size for the
sample is comparable to the range for local galaxies. This rules
out the possibility that the GEMS and GOODs rings studied here are
circumnuclear with the surrounding disks below the detection
threshold.

Integrated restframe (U-B) and (B-V) colors are shown in Figure
\ref{fig:ringsf6}. The chain and bent chain galaxies in the GEMS
field have colors similar to each other, and these are $\sim0.5$
mag bluer than the ring and partial ring galaxies, suggesting more
recent star formation in the chains and bent chains. The colors of
the ring and partial ring galaxies are typical for local galaxies
with intermediate to late types (e.g., Fukugita, Shimasaku, \&
Ichikawa 1995), but the chain and bent chain galaxies are bluer
than local late-type galaxies. These blue colors are consistent
with the common observation that high redshift galaxies of all
types tend to be starbursting (Coe et al. 2006; Elmegreen et al.
2006; de Mello et al. 2006).

The radial profiles in V$_{606}$ of the ring and partial ring
galaxies were found to be approximately exponential in the inner
regions out to the ring, suggesting the underlying disks are the
same as in local spirals. Inner exponentials also occur in local
collisional ring galaxies. The (V$_{606}$ - z$_{850}$) colors of
our rings are 0.3 mag bluer than their average inner disks,
suggesting enhanced ring star formation. Such blue colors are
consistent with observations and models of the Cartwheel galaxy
(Korchagin, Mayya, \& Vorobyov 2001), VII Zw 466 (Thompson \&
Theys 1978), and other collisional ring galaxies (Appleton \&
Marston 1997). These local galaxies were observed in restframe B-V
and V-K and not U-B, as in the present sample, but the relative
blueness of the rings is the same for each.

In contrast, the bent chains generally have little emission
interior to their arcs, as evident in Figure \ref{fig:contour2}.
Some bent chains have stray clumps nearby with approximately the
same colors as the galaxies (e.g., galaxy No. 11357 in COMBO17,
which is the bent chain in Fig.\ref{fig:bentchain} with redshift
0.136, and No. 3618 [z=0.764]). A stray clump with the same
redshift could be a collision partner or it could be a remote part
of the same disk that contains the bent chain. Some chains also
have an asymmetry in the sharpness of their edges on the inner and
outer arcs (e.g. COMBO17 No. 3618 [z=0.764] and 41096 [z=0.113]),
suggesting a disk or partial disk with the bent chain on the rim.
Thus, the bent chains could be partial ring galaxies, but the
galaxy disks around the bent chains would have to be unusually
faint. On the other hand, bent chains could be warped chain
galaxies. Along the lengths of the bent chains, the {\it pvector}
intensity profiles are similar to those of other chain galaxies in
the UDF, which are irregular with no exponential gradients
(Elmegreen, Elmegreen, \& Sheets 2004; Elmegreen et al. 2005b).

The V$_{606}$ surface brightnesses of the galaxies were also
determined using the National Institutes of Health software
package {\it imagej} (Rasband 1997), which allows measurement
along a curved path. The average surface brightness in the
interclump regions between the clumps is plotted for each galaxy
versus $(1+z)^4$ in Figure \ref{fig:ringsf7}. The intensity at a
fixed restframe passband is expected to decrease from cosmological
effects as the inverse of $\left(1+z\right)^4$, as shown by the
dotted line (e.g., Lubin \& Sandage 2001). With bandshifting in
addition, the intensity could vary in a different way, depending
on the star formation rate; higher rates produce intrinsically
bluer galaxies and less of a decrease with
$\left(1+z\right)^{-4}$.

Figure \ref{fig:ringsf7} suggests that the interclump intensities
for the ring and partial ring galaxies are similar to each other
and decrease slightly with increasing $z$, although not in
proportion to $\left(1+z\right)^{-4}$. The spread in surface
brightness for these galaxies is $\sim1.5$ mag arcsec$^{-2}$ at
each redshift. The bent chain galaxies are slightly different,
having a nearly constant apparent interclump surface brightness,
$\sim24$ mag arcsec$^{-2}$, independent of $\left(1+z\right)^4$
(except for one bent chain at $z=0.296$, which is COMBO17 No.
37539). This difference suggests that the interclump regions of
bent chain galaxies have intrinsic surface brightnesses that
increase rapidly with $z$ as a combined result of bandshifting
into a relatively bright uv wavelength and an increasing star
formation rates with look-back time. The rings and partial rings
have a more constant intrinsic interclump V$_{606}$ surface
brightness, which is dimmed by cosmological effects without as
much offset from star formation.

\subsection{Clump properties}

The clump sizes and separations were measured for each galaxy. The
right-hand side of Figure \ref{fig:ringf10a} shows the clump
diameter versus redshift for the GEMS and GOODS galaxies (the
streaks correspond to integer numbers of pixels). The clump
diameters do not depend noticeably on galaxy type. There is a
correlation between minimum clump size and $z$ because of angular
resolution limits. Figure \ref{fig:ringsf8} shows histograms of
the ratio of the clump diameter to the galaxy diameter. The
histograms peak at $\sim0.08$ galaxy diameters regardless of
galaxy type. The drop at lower relative size is probably from the
resolution limit.

The clump separations are typically 1 to 3 clump diameters, and
about 0.1 to 0.2 times the galaxy diameters. These sizes and
proportions are consistent with the formation of clumps by
gravitational instabilities in chain galaxies and collisional
rings, considering that the ambient Jeans length is about the same
as the clump separation. Collisional ring star formation has been
modeled in detail by Struck-Marcell \& Appleton (1987),
Struck-Marcell \& Higdon (1993), and others. Bent chains have the
same clump sizes and relative separations as straight chains.

Figure \ref{fig:Vbull} shows the apparent V$_{606}$ clump
magnitude as a function of redshift for the GEMS and GOODS ring,
partial ring, chain, and bent chain galaxies (left), and for the
UDF chain and clump cluster galaxies (right). The nuclei in the
ring and partial ring galaxies (diamonds in left panel) have
apparent V$_{606}$ mag of 24.5$\pm$2 mag. Aside from the nuclei,
the brightest GEMS and GOODS clumps in each morphology class are
in the range from 25 to 26 mag, with no systematic trend in
redshift. For the UDF clumps, the brightest are at $\sim26$ mag,
which is about the same. However, the faintest detected clumps are
2 mags brighter in GEMS and GOODS than they are in the UDF as a
result of the brighter detection limit of the GEMS and GOODS
surveys. A more detailed examination of the GEMS and GOODS
galaxies also revealed extremely faint clumps, but they are too
faint to measure (e.g., COMBO17 No. 47074 [z=0.837] has very faint
clumps in Fig. \ref{fig:bullsrings}).

The restframe apparent (left) and absolute (right) B magnitudes of
the clumps are shown as a function of redshift in Figure
\ref{fig:Vchain} for galaxies in each sample. These restframe
apparent magnitudes were found by interpolation between the
observed passbands. For GEMS galaxies with $z$ between 0.39 and
0.95, the restframe blue luminosity was taken to be
$L_{B,rest}=L_{V,606} (0.95-z)/(0.95-0.39) + L_{Z,850}
(z-0.39)/(0.95-0.39)$ for observed $V_{606}$ and $z_{850}$
luminosities. The restframe B magnitude is then $-2.5 \log
L_{B,rest}$.  For GOODS and UDF galaxies, which have 4 observed
passbands, the linear interpolation was done between each pair of
passbands.  We converted the apparent restframe magnitudes to
absolute restframe magnitudes by applying the distance modulus
determined from a $\Lambda$CDM cosmology (Spergel et al. 2003).
For both the apparent and absolute magnitudes, the brightest
clumps are equally bright in all galaxies types. The redshift
trend for absolute magnitude is from the size cutoff: only the
largest clumps can be seen at the highest redshift (Fig.
\ref{fig:ringf10a}) and these are the brightest.

Clump ages and masses were estimated from their $V_{606}-z_{850}$
colors and magnitudes using Bruzual \& Charlot (2003) spectral
models that were redshifted and integrated over the filter
functions of the ACS camera.  We assumed models with a metallicity
$Z=0.008$ and a Chabrier (2003) IMF. Intervening hydrogen
absorption was included, following Madau (1995), and internal dust
was included following Rowan-Robinson (2003), with extinction
curves from Calzetti et al. (2000) and Leitherer et al. (2002).
More details are given in Elmegreen \& Elmegreen (2005).

Figure \ref{fig:VZvz} shows the observed colors as a function of
$z$ for the ring and bent chain clumps in the GEMS and GOODS
fields. The clump (V$_{606}$ - z$_{850}$) colors generally range
from 0 to 1 for all galaxies, with an average of 0.8$\pm$0.6 for
the ring clumps and 0.4$\pm$0.4 for the bent chain clumps. There
is a slight redshift dependence towards redder colors with
increasing $z$ for the ring and partial ring clumps.  The
superposed curves are models from Bruzual \& Charlot, discussed
above. The solid curves assume a clump age of $10^9$ years and 4
different star formation rate models: exponential decays with
timescales of $10^7$ yr (top curve), $3\times10^8$ yr, and $10^9$
yr, and a continuous star formation rate (bottom curve). The
dashed curves assume a clump age of $10^8$ yrs, with 4 similar
star formation rate models, as indicated, and the dotted curve
assumes a clump age of $10^7$ yr and an exponentially decaying
star formation rate with a decay time of $10^7$ yr.

The bent chain clumps are bluer than the ring and partial ring
clumps for the same $z$, suggesting that the bent chain clumps are
slightly younger than the ring clumps.  Most of the bent chain
clumps match models with ages of $\sim10^7$ yrs to $\sim10^8$ yr
for long decay times. Only one bent chain at $z\sim0.7$ may have
clumps as old as $10^9$ yrs. In contrast, clumps in the ring and
partial ring galaxies are best matched by ages of $10^8$ yrs for
various decay times, or $10^9$ years with a long decay time.

From the clump ages and apparent magnitudes, we can estimate
masses from Bruzual \& Charlot (2003) models, as shown in Figure
\ref{fig:Mass}. This figure plots mass versus clump age for a
typical clump apparent magnitude of V$_{606}=27$ (see Fig.
\ref{fig:Vbull}), from which other magnitudes and masses can be
scaled. Each curve is a different exponential decay time in a star
formation model: $10^7$ yr (top curves), $3\times10^7$ yr, $10^8$
yr, $3\times10^8$ yr, $10^9$ yr, and continuous (bottom curves).
Considering the ages and decay times estimated in the previous
paragraph, and the apparent V$_{606}$ magnitudes of the clumps, we
infer that the clump masses average $\sim5\times10^6$ M$_{\odot}$
at low redshift to a few $\times10^8$ M$_{\odot}$ at $z\sim1$. The
clump masses are a factor of 2 or 3 lower for the bent chains than
the ring and partial ring galaxies because, although the clump
magnitudes are about the same, the bent chain clumps have slightly
younger ages. Clump mass increases with redshift, as does the
clump size and absolute magnitude discussed above, because of our
size cutoff.

\subsection{Comparison with Local Ring Galaxies}

\subsubsection{Collisional Ring Galaxies}

Several local collisional ring galaxies have ring morphologies
similar to those in Figures \ref{fig:bullsrings} and
\ref{fig:bullspartial}. Some have off-center nuclei and others are
empty.  Some resonance ring galaxies resemble our ring galaxies
too. Because our GEMS and GOODS galaxies have redshifts averaging
0.78$\pm$0.33, the restframe is $ \sim$ $3400\AA$ for the
V$_{606}$ images and $4800\AA$ for the z$_{850}$ images. To make
appropriate comparisons, we measured clumps in local galaxies from
UV or B-band archival images on the website
nedwww.ipac.caltech.edu, as available. Some of these images are at
high resolution with HST, while others are from ground-based
Schmidt or 5-m telescopes. The following nearby collisional ring
galaxies were compared to our GEMS and GOODS sample: Arp 107, 146,
147, 148, 149; NGC 922; Cartwheel; AM0644-74;  and VII Zw 466
(+UGC07683). For all of them, clump diameters ranged from 0.2 to 7
kpc (the latter with poor resolution), with a peak in the size
distribution at 1.4 $\pm$ 1.8 kpc. The ratio of clump size to ring
size was the same as for all the GEMS and GOODS samples, again
peaking at about 0.1 (see Fig. \ref{fig:ringsf8}).

The clumps in local collisional ring galaxies are bluer than the
inner disks, as are the clumps in the GEMS and GOODS ring
galaxies.   For example, an optical study of several ring galaxies
by Theys \& Spiegel (1976) indicated ring colors similar to the
blue colors of Magellanic-type irregulars. Analysis of the colors
of 11 ring galaxies by Appleton \& Marston (1997) showed that all
of them have an outward blue gradient. Optical colors of clumps in
VII Zw 466 suggested ages of $\sim$ 10$^8$ yrs, consistent with
the inferred ring interaction age (Thompson \& Theys 1978). Three
of the star forming clumps in the ring of the Cartwheel galaxy
(A0035=Arp10) have absolute B magnitudes of -17 to -18 mag,
similar to the restframe absolute B clump magnitudes found here
(Fig. \ref{fig:Vchain}); the total stellar mass associated with
the HII regions is about $10^8$ $M_{\odot}$ (Fosbury \& Hawarden
1977).  A similar object is the large southern ring galaxy AM
0644-741. It has massive star formation apparently triggered in a
double ring by a close encounter (Arp \& Madore 1987). Higdon \&
Wallin (1997) used H$\alpha$ measurements to infer star formation
rates in AM 0644-741 and found a total new stellar mass of several
times $10^8$ $M_{\odot}$ in the ring, considering the ring age of
110 Myr.  These ages and masses are all similar to what we found
for distant ring clumps.

Some local collisional ring galaxies show spoke-like features that
connect the inner cores to the rings, such as the Cartwheel and
NGC 922. Although such features are difficult to discern in our
distant galaxy sample, the top middle galaxy in Figure
\ref{fig:bullspartial} (COMBO17 No. 47074 [z=0.837]) shows two
inner arcs. Some of the others show hints of substructure also.
Non-axisymmetric features like these may grow from fragmentation
and collapse of material in outwardly propagating waves (Hernquist
\& Weil 1993; Struck 1997).  A bar-like structure is seen in the
top right galaxy in Figure \ref{fig:bullspartial} (No. 48709,
$z=0.494$). Huang \& Stewart (1988) showed that a near-planar
collision by an intruder can produce such a bar.

Several UDF clump cluster galaxies resemble local ring galaxies
because they have massive clumps in a circle around an empty
center, as noted by Wong et al. (2006). Most clump-clusters have
irregularly placed clumps, though, so the fraction of these
galaxies that might have collision or resonance rings is probably
small.

The 25 galaxies identified as collisional ring galaxies in
parallel fields out to z=1 with WFPC2 F814W observations (Lavery
et al. 2004) show many similarities to our GEMS and GOODS ring
galaxies. Galaxies numbered  2, 8, 10, 11, 12, and 17 in their
Figure 1 have symmetric rings and centered nuclei, like some of
ours in Figure \ref{fig:bullsrings}. The remainder of the Lavery
et al. sample match better the partial ring galaxies in Figure
\ref{fig:bullspartial}.

\subsubsection{Resonance Ring Galaxies}

\begin{figure}\epsscale{1.0}
\plotone{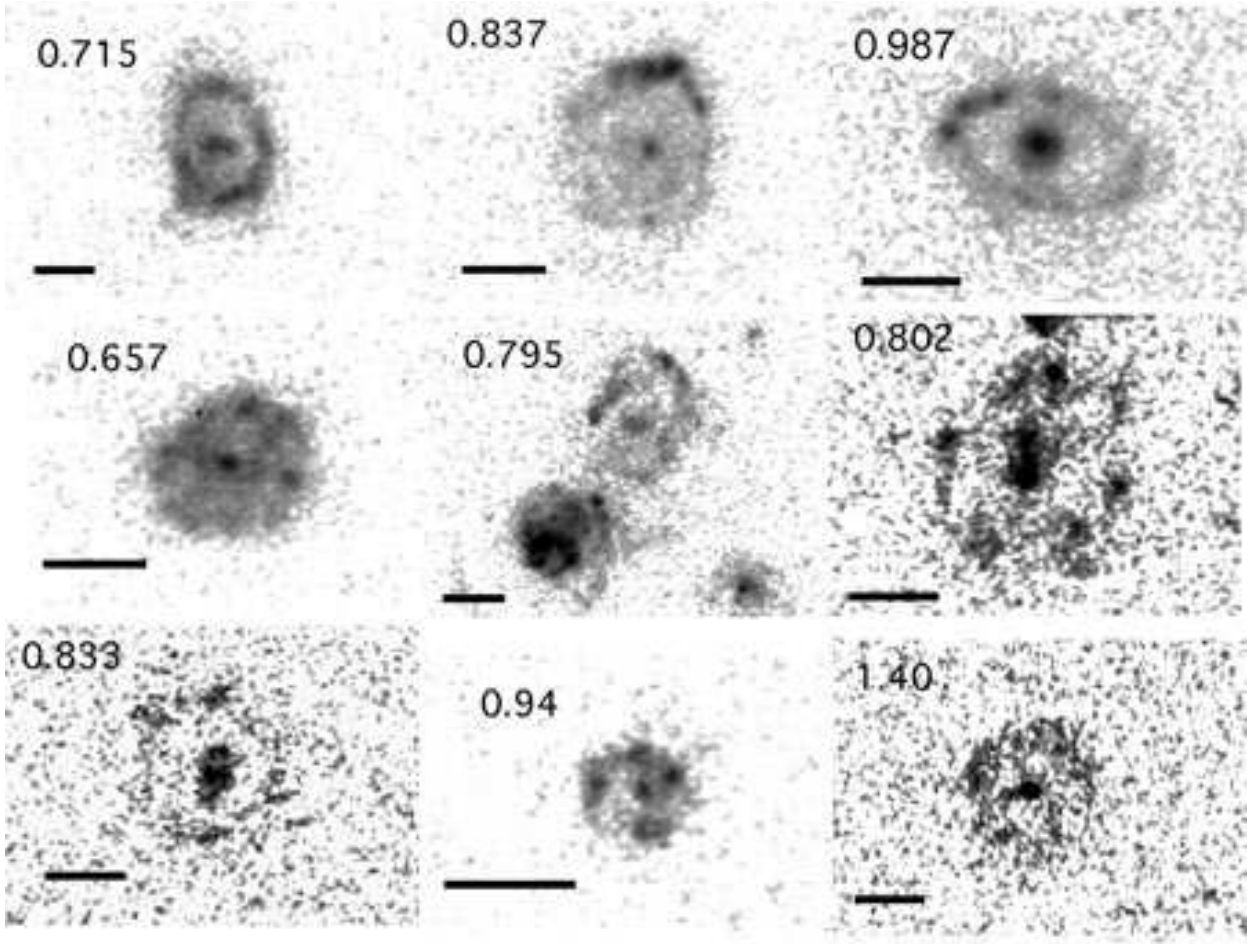}\caption{$V_{606}$-band images of ring galaxies in
the GEMS and GOODS fields. Redshifts are in the upper left corner;
horizontal lines represent 1''. (Image degraded for
astroph.)}\label{fig:bullsrings}\end{figure}

\begin{figure}\epsscale{1.0}
\plotone{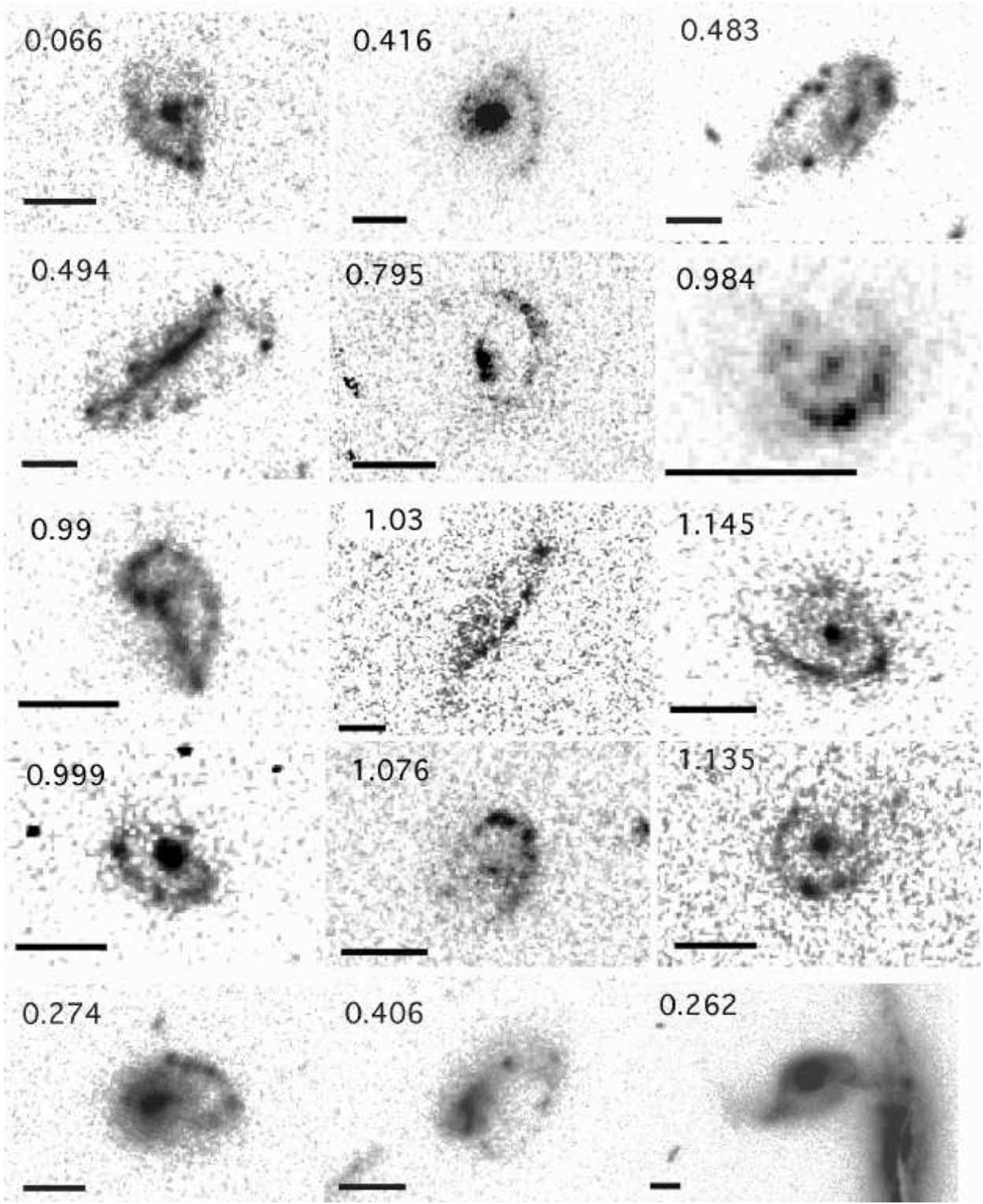}\caption{$V_{606}$-band images of partial ring
galaxies in the GEMS and GOODS fields.  Redshifts are in the upper
left corner; horizontal lines represent
1''.}\label{fig:bullspartial}\end{figure}

\begin{figure}\epsscale{1.0}
\plotone{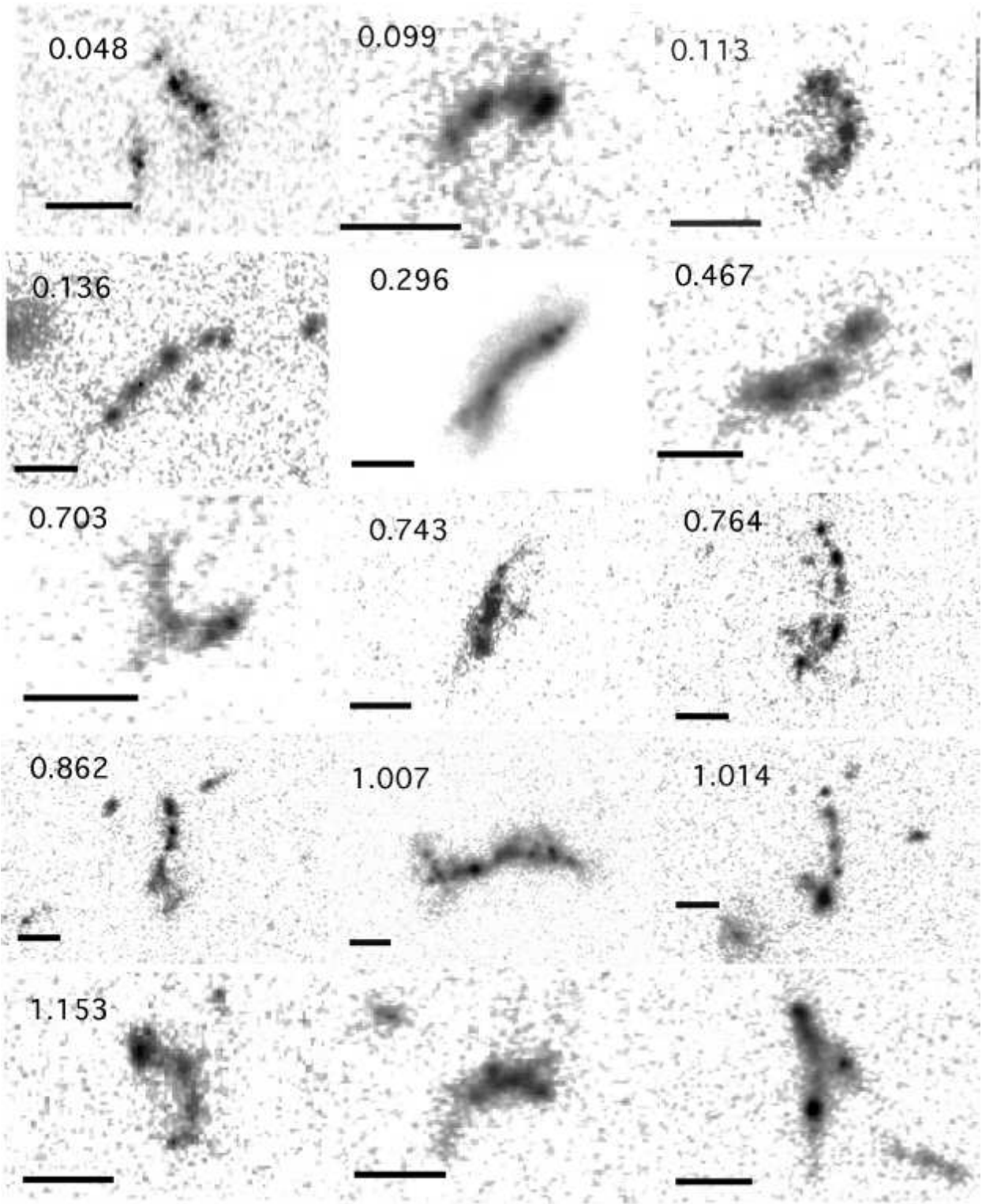}\caption{$V_{606}$-band images of bent chain
galaxies in the GEMS and GOODS fields. Redshifts are in the upper
left corner; horizontal lines represent
1''.}\label{fig:bentchain}\end{figure}

\begin{figure}\epsscale{1.0}
\plotone{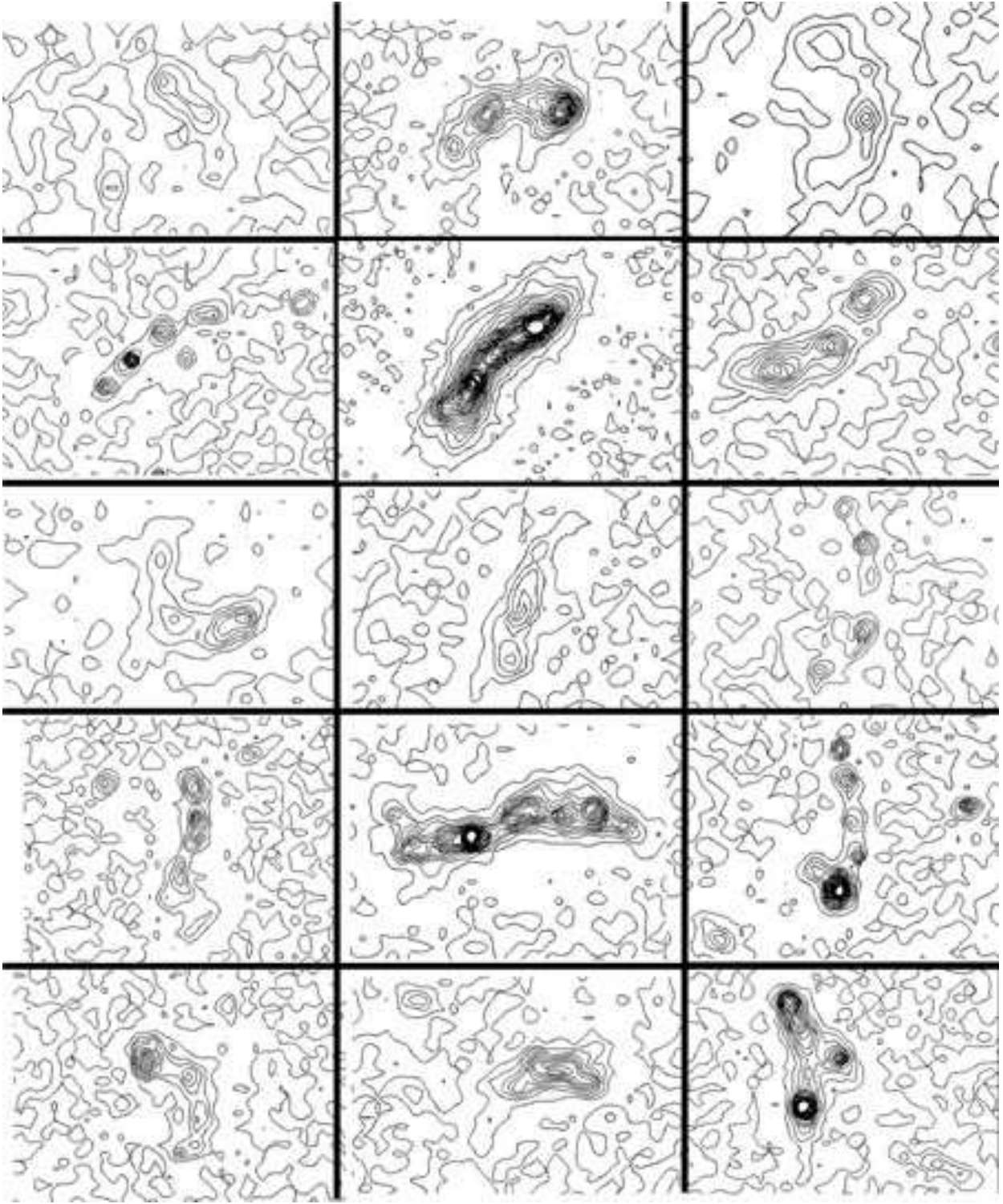}\caption{2-$\sigma$ contours of the bent chain
galaxies in the previous figure. The lowest contour (2-$\sigma$)
is 25 mag arcsec$^{-2}$. There is no face-on component of a disk
inside the arcs, suggesting either that the underlying disks are
unusually faint or these objects are unlike collisional or
resonance ring galaxies. (Image degraded for astroph.)
}\label{fig:contour2}\end{figure}

\begin{figure}\epsscale{1.0}
\plotone{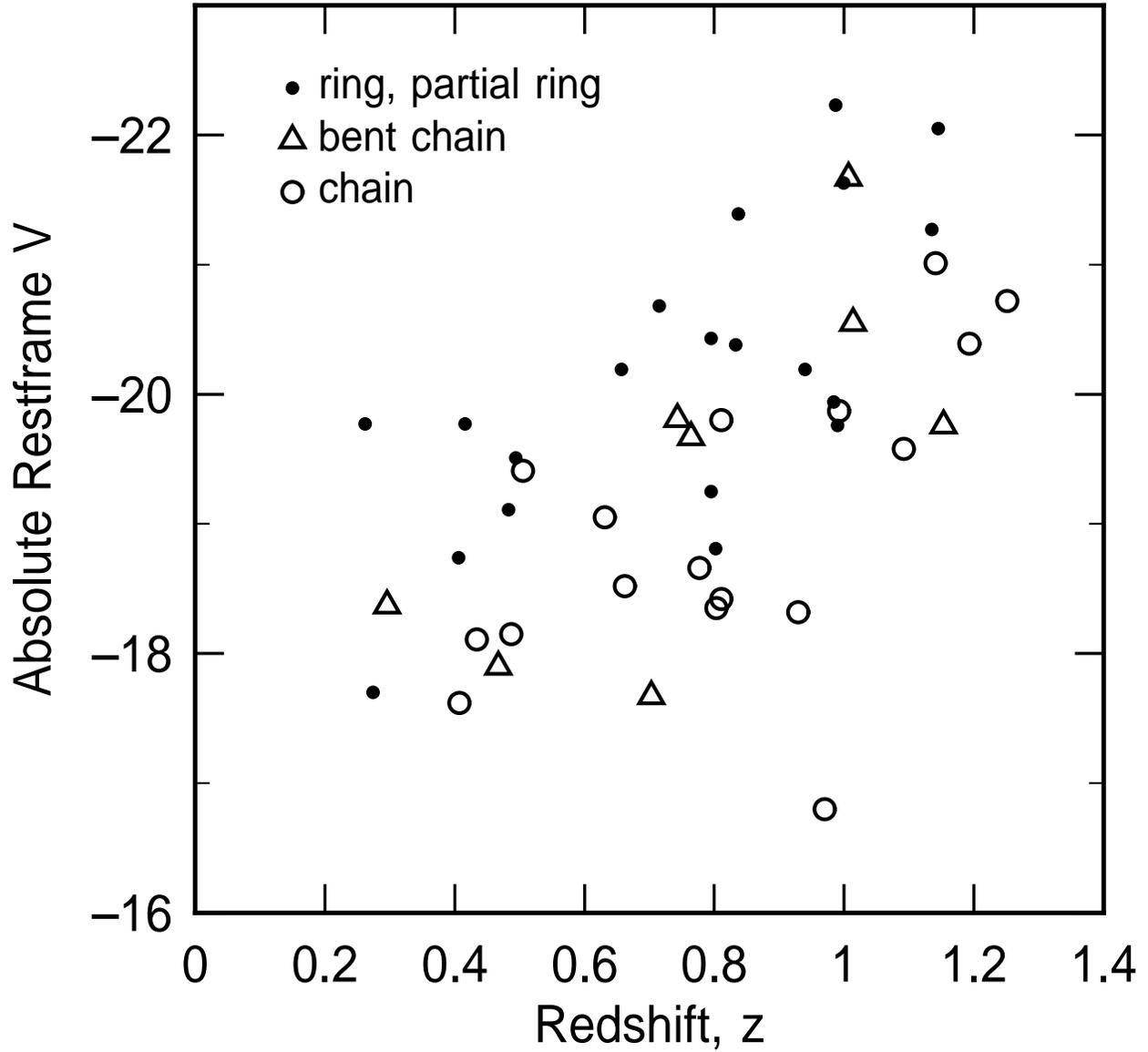}\caption{Absolute restframe V-band integrated
magnitudes for ring and partial ring, bent chain, and chain
galaxies in the GEMS and GOODS fields. The ring and partial ring
galaxies are brighter on average than the chains and bent chains
at a given redshift.}\label{fig:ringsf5}\end{figure}

\begin{figure}\epsscale{1.0}
\plotone{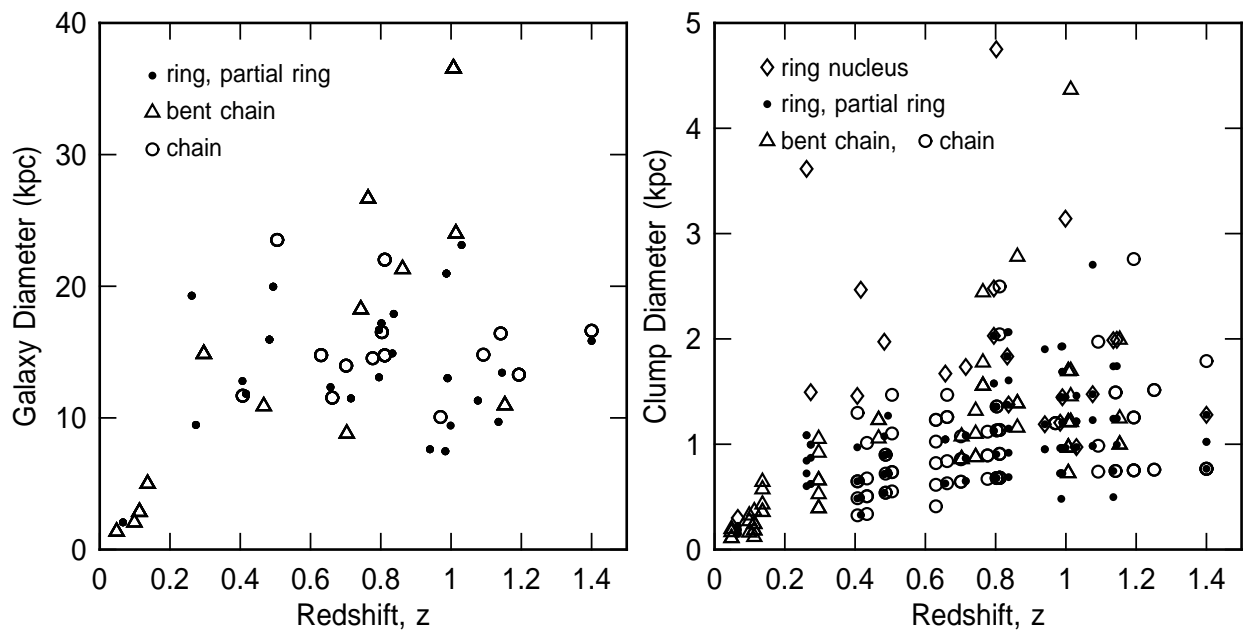}\caption{(left) Diameters of surveyed galaxies in
the GEMS and GOODS fields. Above $z\sim0.2$ there is no obvious
correlation with redshift for any type. (right) Diameters of the
clumps in the GEMS and GOODS galaxies, versus redshift. The band
structure is from integer values of pixel size. The correlation is
the result of decreasing spatial resolution with increasing $z$.
There is no obvious difference in size for the various types. The
ring galaxy nuclei (diamonds) are generally larger than the
clumps.}\label{fig:ringf10a}\end{figure}

\begin{figure}\epsscale{1.0}
\plotone{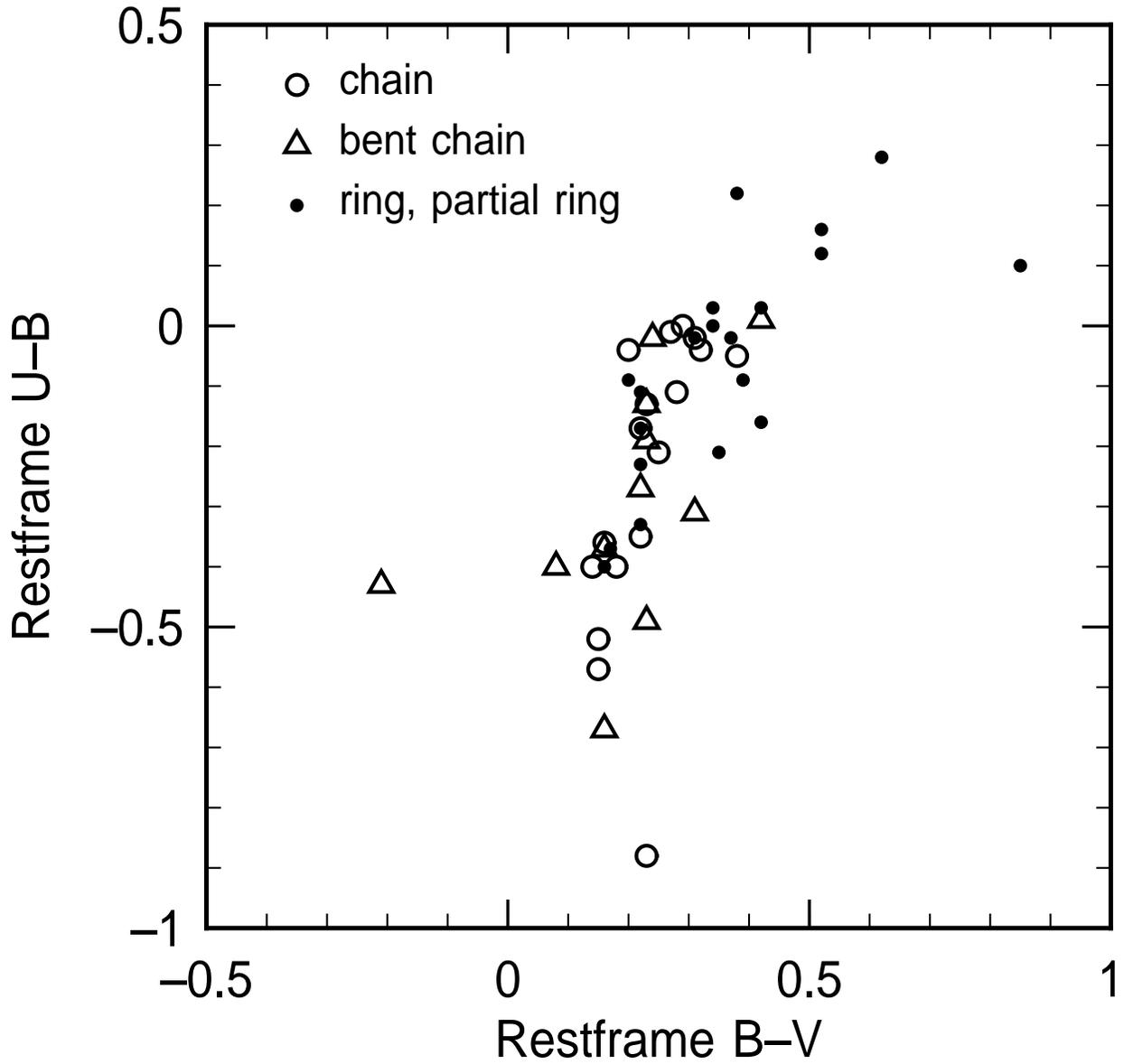}\caption{Restframe (U-B) and (B-V) integrated
colors for ring and partial ring, bent chain, and chain galaxies
in the GEMS and GOODS fields. The chain and bent chain galaxies
are bluer than the ring and partial ring galaxies, suggesting
younger average ages.}\label{fig:ringsf6}\end{figure}

\begin{figure}\epsscale{1.0}
\plotone{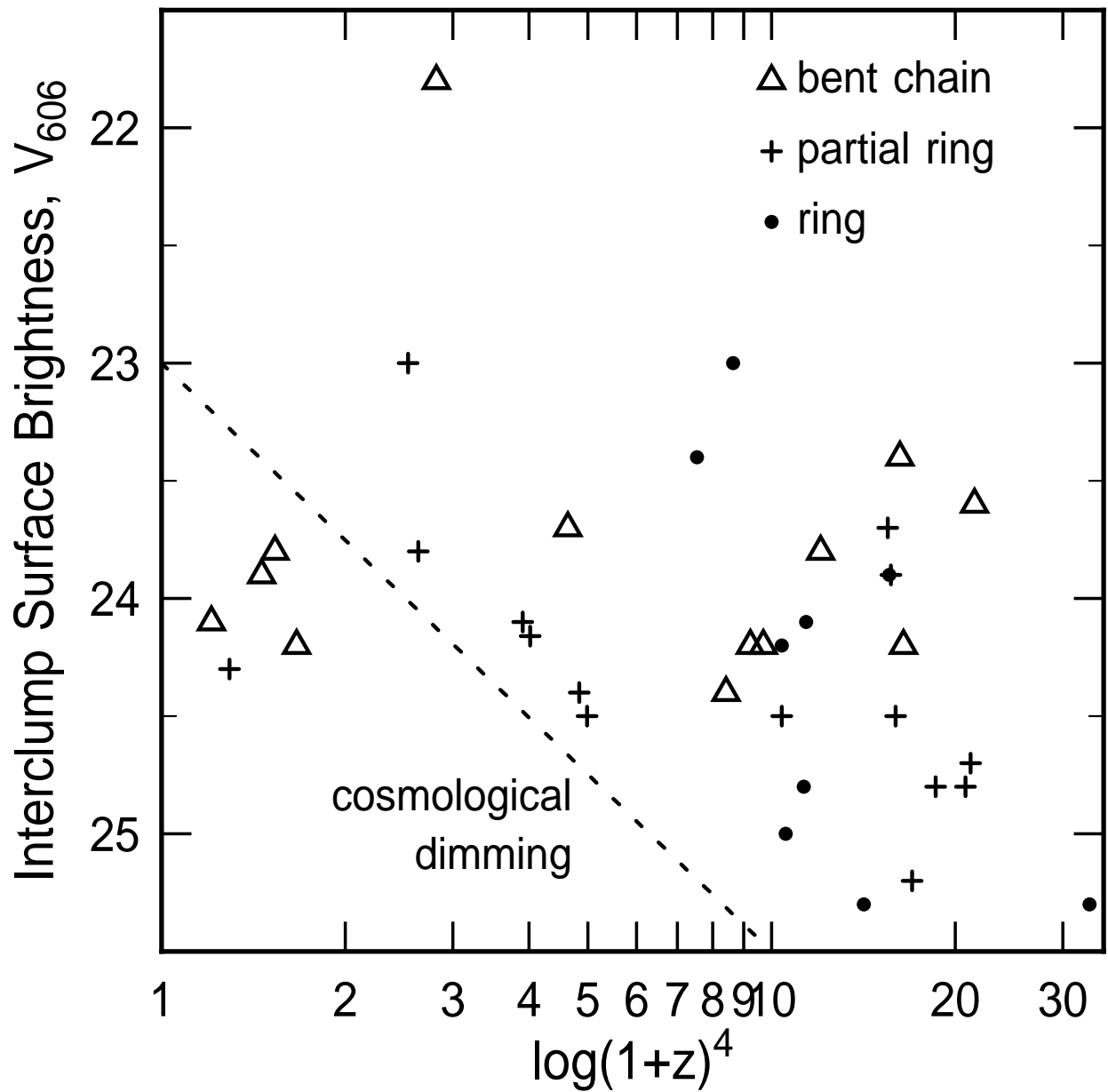} \caption{Apparent V$_{606}$ surface brightnesses
in mag arcsec$^{-2}$ of the interclump regions as a function of
$(1+z)^4$. The bent chains have a more constant V$_{606}$
interclump surface brightness than the rings and partial rings,
suggesting a difference in their average star formation
properties.} \label{fig:ringsf7}\end{figure}

\begin{figure}\epsscale{1.0}
\plotone{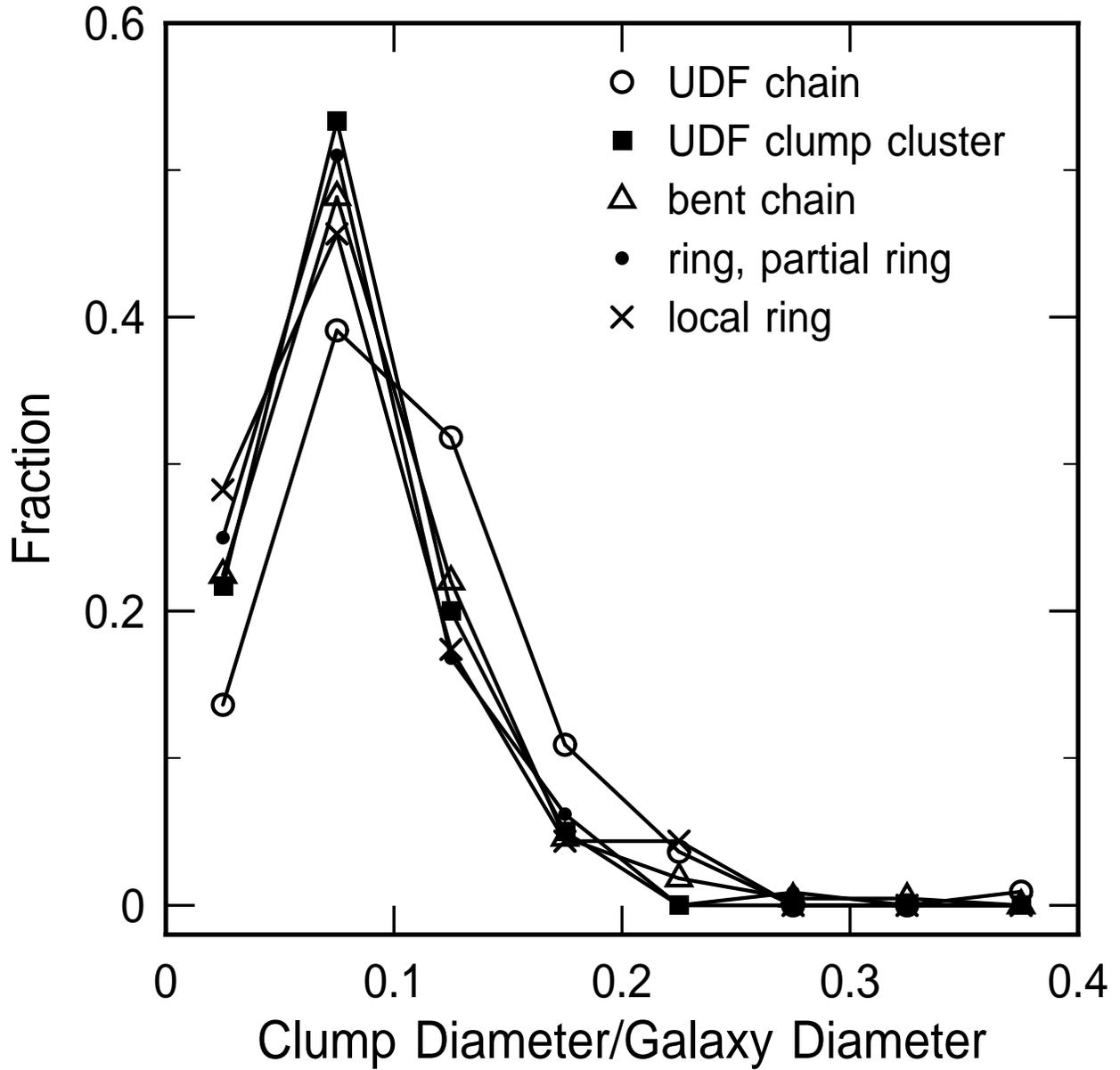} \caption{The clump sizes relative to the galaxy
sizes are shown for each type, including local ring galaxies. The
relative sizes are comparable to or larger than $0.08$.}
\label{fig:ringsf8}\end{figure}

\begin{figure}\epsscale{1.0}
\plotone{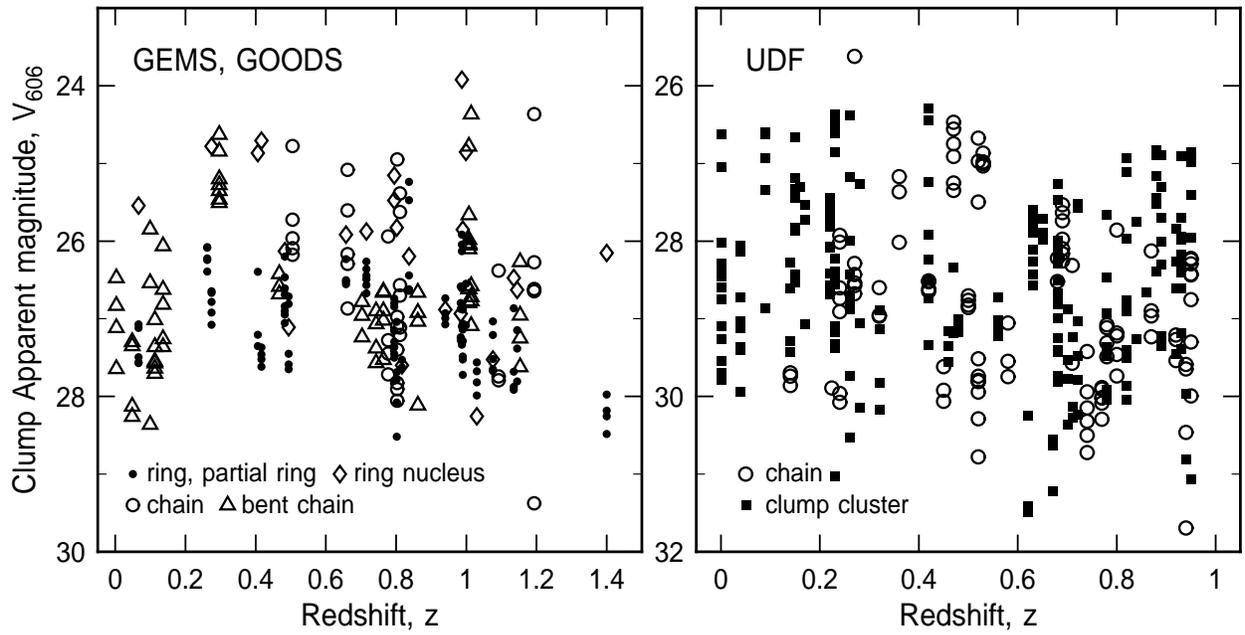}\caption{Observed V$_{606}$ mag for nuclei and
clumps in GEMS and GOODS ring, partial ring, chain, and bent chain
galaxies (left) and for clumps in UDF clump-cluster and chain
galaxies (right) as functions of redshift.}
\label{fig:Vbull}\end{figure}

\begin{figure}\epsscale{1.0}
\plotone{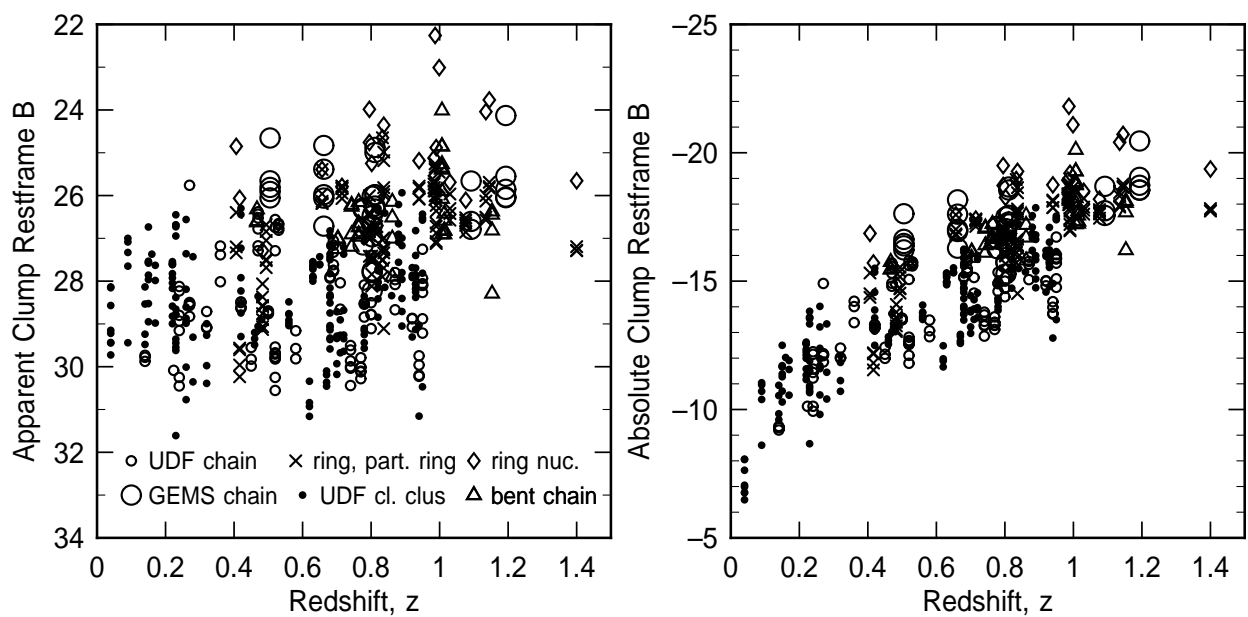} \caption{Rest-frame B-band apparent (left) and
absolute (right) clump magnitudes versus redshift for GEMS, GOODS,
and UDF galaxies in the sample. The brightest clumps are equally
bright in chain, clump cluster, bent chain, and ring galaxies.
Nuclei in ring and partial ring galaxies are indicated by
diamonds.} \label{fig:Vchain}\end{figure}

\begin{figure}\epsscale{1.0}
\plotone{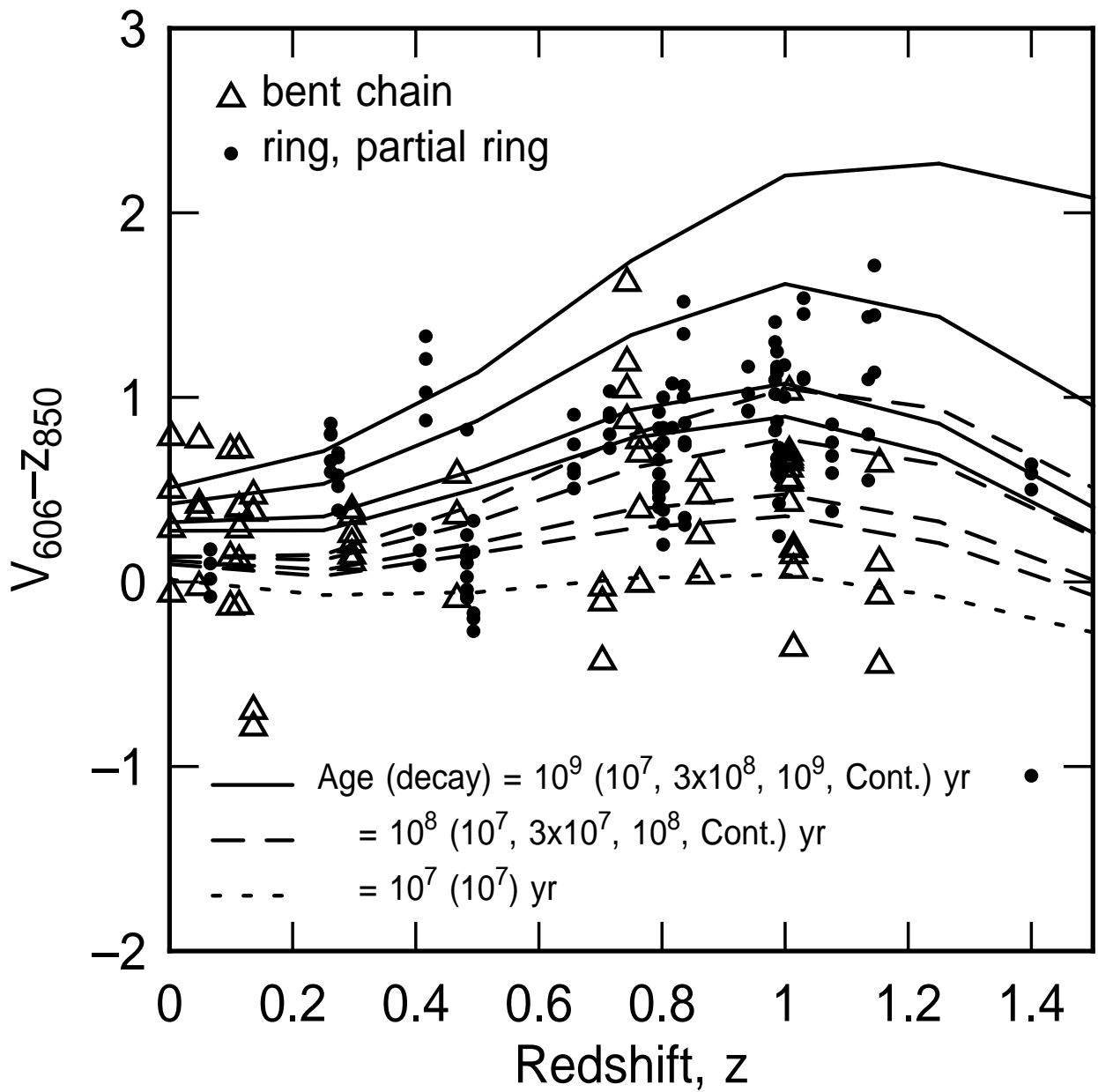} \caption{Observed (V$_{606}$-z$_{850}$) color
versus redshift $z$ for ring and bent chain galaxy clumps in the
GEMS and GOODS fields. Curves are models from Bruzual \& Charlot
(2003) with various clump ages and star formation decay times.}
\label{fig:VZvz}\end{figure}

\begin{figure}\epsscale{1.0}
\plotone{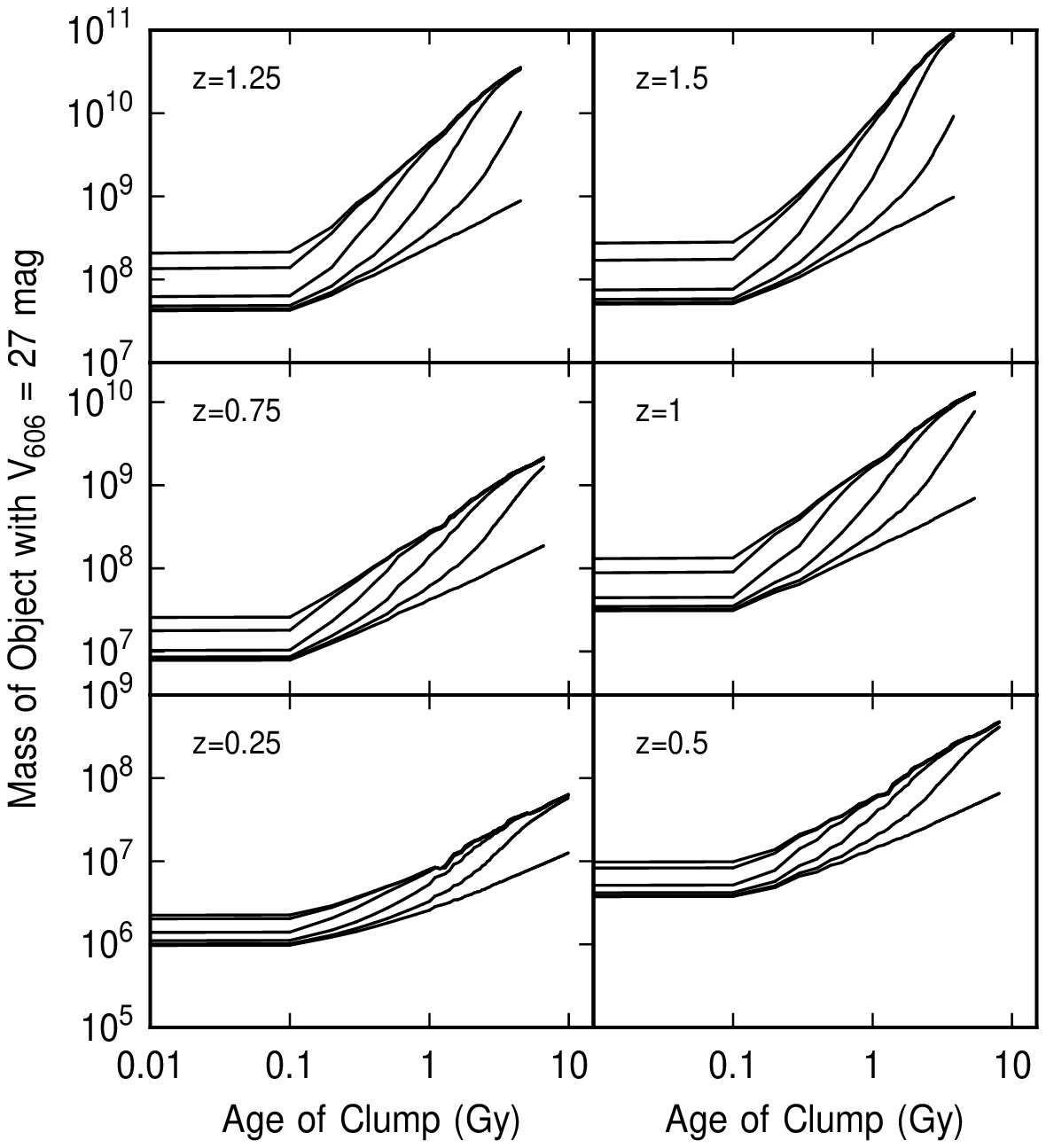} \caption{Expected mass as a function of star
formation duration for a clump with an observed V$_{606}$ mag of
27 for 6 different redshift bins. Star formation duration was
estimated from a comparison of the previous two figures.}
\label{fig:Mass}\end{figure}

Several of the galaxies in Figure \ref{fig:bullsrings} have
symmetric rings with centered nuclei, making them look somewhat
like local galaxies with strong outer resonance rings. For
comparison, we examined the local outer-ring galaxies IC 1993, NGC
6782, NGC 1433, IC 1438, and NGC 6753. The local rings are similar
in size to the GEMS and GOODS rings, but the local star formation
clumps in the rings are smaller and much fainter. The clumps in
local resonance rings have average U and B-band sizes equal to
only $\sim0.04$ times the galaxy diameters, whereas the GEMS and
GOODS clumps were typically comparable to or larger than 0.08
galaxy diameters (Fig. \ref{fig:ringsf8}). Clump sizes that are as
small as in local ring galaxies might not be resolved in the GEMS
and GOODS fields, but still, the largest ring clumps at high
redshift are larger than the largest ring clumps locally. This
difference could be the result of a higher star formation rate in
higher redshift galaxies. The local outer-resonance ring galaxies
tend to be early-type spirals, which have low star formation rates
today.

Because bars can produce resonance rings and bars are red, some of
the GEMS and GOODS ring galaxies might have bars that cannot be
seen at visible wavebands for our average $z\sim0.78\pm0.33$. A
mid-UV study of local galaxies (Windhorst et al. 2002) reveals
that morphology is often similar from UV to optical, with a bar
showing up in B as well as UV. Sometimes the change with passband
is dramatic, however; in NGC 6782 the optical bar is visible out
to B band but disappears in the UV. While such a change is
possible for our highest redshift galaxies, even at z=0.93 we
observe the restframe B band in the z$_{850}$ images and most bars
should still be present there. Only 7 of our ring and partial ring
galaxies are at higher redshift than this. Thus we looked for bars
in our ring sample. A few of the galaxies in Figure
\ref{fig:bullsrings} (top left [No. 53346, $z=0.715$], middle
right [No. 49092, $z=0.802$], bottom left [No. 43780, $z=0.833$],
bottom right [No. 62696, $z=1.40$]) and Figure
\ref{fig:bullspartial} (top middle [No. 34409, $z=0.483$], top
right [No. 48709, $z=0.494$]) show slight elongations in their
centers that could be bars or oval distortions capable of
producing an outer resonance ring. None of these bars is as
prominent as the strong bars seen in today's SB0 ring galaxies
(e.g. Buta \& Combes 1996).

Collisional ring galaxies generally have small intruder galaxies
nearby, so we examined the GEMS and GOODS fields surrounding the
symmetric cases for possible companions. We found that four of the
nine ring galaxies in Figure \ref{fig:bullsrings} (top left [No.
53346, $z=0.715$], middle-middle [No. 50905, $z=0.795$],
bottom-left [No. 43780, $z=0.833$], bottom-right [No. 62696,
$z=1.40$]) have small companions about 10 px in diameter with
redshifts differing from the ring galaxy by less than 0.1 and
appearing within 1 to 3 galaxy diameters. These four cases could
therefore have collisional rings and not resonance rings.

The galaxies without clear companions (Nos. 47074, 44999, 58535,
49092, 25076) are the most likely examples of resonance ring
galaxies. Still, the morphology of these five candidates differs
from the morphology of local outer-resonance ring galaxies. Only
one of the five has an elongated nucleus that could be a bar (No.
49092 [$z=0.802$]); the rest have small circular nuclei. This
contrasts with local outer-ring galaxies, which are mostly barred
or have strong spiral arms. In de Vaucouleurs \& Buta (1980),
there are 46 galaxies with outer rings, and 17 with the A-type bar
class (nominally, no bar). Of these, 4 looked barred after all, 1
is a strong merger with long tidal arms, 4 have flocculent arms
throughout that are somewhat brighter in the ring region (but not
a real ring), 2 others have spiral arms that wrap to an incomplete
ring, and 2 have no visible ring (based on B images in
nedwww.ipac.caltech.edu). The remaining three (NGC 1068, NGC 4736,
NGC 3977) have smooth weak outer rings and large bright bulges and
inner disks.  These three local galaxies are the most similar to
our four non-barred outer-ring candidates, although they are not
identical.  Also, the bar fraction for the best examples of local
outer ring galaxies is very high, while it is only 20\% in our
sample.  It is conceivable that our four non-barred cases had bars
when they formed their rings and then the bars dissolved later.
Such ring formation had to take $\sim6$ Gy or less for galaxies at
$z\sim0.8$. Why most of the local outer-resonance ring galaxies
still have their bars when the distant ones do not is a mystery.

The average major axis diameter of the 5 candidate resonance rings
is $15\pm5$ kpc. The average diameter of the outer rings in 46
local galaxies in the ring catalog of de Vaucouleurs \& Buta
(1980) is $20\pm10$ kpc (using a Hubble constant of 71 km s$^{-1}$
Mpc$^{-1}$).  The local and high redshift diameters are
sufficiently close to each other that there could not have been a
major change in outer ring position over the last $\sim6.6$ Gy
since $z=0.78$. The slight increase seen here is consistent with
an overall growth of galaxies over this time, by $\sim30$\% or so
(with large uncertainties), or with a comparable slow down in the
pattern speed for whatever drives the resonance.

Bar pattern speeds are expected to slow down as the bar transfers
angular momentum to the halo (Athanassoula 2003). Athanassoula
(2003) found that this interaction can slow down a bar in the last
$\sim6$ Gy by 10\% to 40\%, depending on the relative mass of the
disk, bulge, and halo (see her figures 10 and 11). This range is
consistent with the small increase in ring diameter found here for
the same period of look-back time.

\subsubsection{Accretion Rings}

Other local analogs for the GEMS and GOODS galaxies are accretion
rings (see Buta \& Combes 1996), such as Hoag's Object, A1515+2146
(Schweizer et al. 1987), which either has a central E0 galaxy that
accreted gas, or was once barred. Other accretion rings include
the more commonly described polar ring galaxies, which tend to
have S0 galaxies in their centers and rings perpendicular to the
disks (Whitmore et al. 1990). To match our observed galaxies, we
would have to be viewing such a galaxy edge-on so the ring looks
circular, but we do not see any extended central component that
could be an edge-on spiral disk.  Also, the atlas by Whitmore et
al. shows that polar rings tend to be much smoother than the
clumpy rings of the GEMS and GOODS galaxies.

\section{Conclusions}

A total of 9 ring and 15 partial ring galaxies was found through
visual examination of the GEMS and GOODS fields. An additional 15
bent chain galaxies was also observed. The clumps in the rings
have diameters of $\sim$1 kpc, depending on redshift. This is
similar to the diameters of clumps in local collisional ring
galaxies. The bent chains have restframe colors that are bluer
than those in ring and partial ring galaxies.  The individual
clumps in the bent chain galaxies are also bluer than the clumps
in rings and partial rings. The average star formation ages of the
clumps in bent chains equals 10$^7$ to a few times $10^8$ yr; in
the rings and partial rings it is a few times $10^8$ to $10^9$
yrs. Clump masses range from a few x 10$^6$ to a few x 10$^8$
M$_{\odot}$, with slightly lower values in bent chains. The
measured clump mass increases with redshift because of resolution
limits.

The morphologies and dimensions of local collisional ring galaxies
are similar to the morphologies and dimensions of many of the ring
galaxies in the GOODS and GEMS fields.  Their clumps also have
about the same sizes, ages, and masses. This similarity implies
that many of the GEMS and GOODS rings and partial rings formed by
plunging impacts of small companions through pre-existing spiral
galaxies.

Other galaxies in our sample, particularly the most isolated and
symmetric and with the most centered nuclei, could have resonance
rings instead of collisional rings.  They differ from local
resonance ring galaxies in having more star formation at high
redshift, and therefore larger and bluer clumps in the rings. Some
have slight indications of a bar, although the bar fraction seems
lower for the best resonance ring cases than it is in the local
Universe. The existence of outer resonance rings in GOODS and GEMS
galaxies implies it takes only $\sim6$ Gy for the rings to form.
After aging for another $\sim7$ Gy and avoiding disruptive
collisions, they could turn into the modern versions of
outer-resonance rings, which are smooth, red, and usually in early
type galaxies. If this is the case, then the similarity in
diameter between the rings in our sample and the local outer
resonance rings (local rings are $\sim30$\% larger) implies that
spiral and bar pattern speeds have not changed by more than a
factor of $\sim1.3$ during last half of the life of the galaxy.
This result may have important implications for nuclear gas
accretion, which can speed up a bar, and bar-halo interactions,
which can slow down a bar (Athanassoula, Lambert \& Dehnen 2005).
For example, bar-halo interactions studied by Athanassoula (2003)
slow down a bar and extend the outer Lindblad resonance by about
this much during the same time interval.

The bent chain galaxies in our survey differ from the ring
galaxies, having younger, slightly less-massive clumps and no
evidence for an underlying, nearly face-on disk. The bent chains
could be edge-on, clumpy, and bulge-free disks that got warped by
an interaction. They resemble the straight chain galaxies that are
observed in deep fields. Local disk galaxies do not have such
large clumps as chains, nor are the local disks warped into
similar arc-like shapes.

D.M.E. thanks the staff at Space Telescope Science Institute for
their hospitality during her stay as a Caroline Herschel Visitor
in October 2005. We thank the referee for useful suggestions about
resonance ring galaxies.

\clearpage

\begin{table}
\footnotesize
\begin{center}
\caption{Ring and Bent Chain Galaxies\label{tab:rings}}
\end{center}
\begin{tabular}{ccc}
\tableline\tableline

catalog no. &   redshift z  &   diameter (kpc)  \\
\tableline
rings   &       &       \\
53346   &   0.715   &   11.5    \\
47074\tablenotemark{a}   &   0.837   &   17.9    \\
44999\tablenotemark{a}   &   0.987   &   20.9    \\
58535\tablenotemark{a}   &   0.657   &   12.3    \\
50905   &   0.795   &   16.6    \\
49092\tablenotemark{a}   &   0.802   &   17.1    \\
43780   &   0.833   &   14.9    \\
25076\tablenotemark{a}   &   0.94    &   7.6 \\
62696   &   1.40    &   15.8    \\
\tableline
partial rings   &       &       \\
25065   &   0.066   &   2.0 \\
36857   &   0.416   &   11.8 \\
34409   &   0.483   &   16.0    \\
48709   &   0.494   &   20.0    \\
14373   &   0.795   &   13.1   \\
21605   &   0.984   &   7.5   \\
38657   &   0.99    &   13.0    \\
34474   &   1.03    &   23.1    \\
41925   &   1.145   &   13.4    \\
34244   &   0.999   &   9.4 \\
23459   &   1.076   &   11.3    \\
36947   &   1.135   &   9.7 \\
21635   &   0.274   &   9.5 \\
29474   &   0.406   &   12.8    \\
25874   &   0.262   &   19.4    \\
\tableline
bent chains\tablenotemark{b} &       &       \\
26239   &   0.048   &   1.4 \\
27211   &   0.099   &   2.1 \\
41096   &   0.113   &   2.9 \\
11357   &   0.136   &   4.9 \\
37539   &   0.296   &   14.9    \\
30656   &   0.467   &   10.9    \\
12329   &   0.703   &   8.8 \\
48146   &   0.743   &   18.2    \\
3618    &   0.764   &   26.6    \\
23308   &   0.862   &   21.3    \\
34857   &   1.007   &   36.5    \\
54277   &   1.014   &   24.0    \\
28154   &   1.153   &   10.9     \\
\end{tabular}
\tablenotetext{a}{These galaxies have no companions 10 px or
smaller in diameter that are within 3 galaxy diameters projected
distance and within a redshift of 0.1. They are symmetric with
centralized nuclei, so they are candidates for resonance ring
galaxies. One of these, 49092, has a nucleus that appears slightly
elongated and could be a bar.} \tablenotetext{b}{The last two
galaxies in Fig. 3 are not listed in the COMBO17 catalog.}
\end{table}

 \clearpage
 \clearpage
 \clearpage
 \clearpage

 \clearpage

 \clearpage

 \clearpage
 \clearpage
 \clearpage

 \clearpage

 \clearpage

 \clearpage

 \clearpage


\begin{thebibliography}{}

\bibitem[]{467} Abraham, R., Tanvir, N., Santiago, B., Ellis, R.,
Glazebrook, K., \& van den Bergh, S. 1996b, MNRAS, 279, L47

\bibitem[]{470} Abraham, R., van den Bergh, S., Glazebrook, K.,
Ellis, R., Santiago, B., Surma, P., \& Griffiths, R. 1996a, ApJS,
107, 1

\bibitem[]{474} Appleton, P., Charmandaris, V., \& Struck, C. 1996,
ApJ, 468, 532

\bibitem[]{477} Appleton, P., \& Marston, A. 1997, AJ, 113, 201

\bibitem[]{479} Appleton, P., \& Struck-Marcell, C. 1987, ApJ, 318,
103

\bibitem[]{482} Arp, H. 1966, Atlas of Peculiar galaxies (Pasadena:
California Inst. of Technology)

\bibitem[]{485} Arp, H.J., \ Madore, B.F. 1987, A Catalog of
Southern Peculiar Galaxies and Associations (Cambridge: Cambridge
Univ. Press)

\bibitem[]{} Athanassoula, E. 2003, MNRAS, 341, 1179

\bibitem[]{489} Athanassoula, E., \& Bosma, A. 1985, ARAA, 23, 147

\bibitem[]{} Athanassoula, E., Lambert, J.C., \& Dehnen, W. 2005, MNRAS, 363, 496

\bibitem[]{491} Barnes, J., \& Hernquist, L. 1992, ARAA, 30, 705

\bibitem[]{493} Beckwith, S., et al. 2004, Hubble Ultra Deep Field
Online Catalog, Space Telescope Science Institute,
http://archive.stsci.edu

\bibitem[]{497}Benitez, N. 2000, ApJ, 536, 571

\bibitem[]{806} Bruzual, G. \& Charlot, S. 2003, MNRAS, 344, 1000

\bibitem[]{} Buta, R., \& Combes, F. 1996, Fund. Cosmic Physics,
17, 95

\bibitem[]{813} Calzetti, D., Armus, L., Bohlin, R.C., Kinney,
A.L., Koornneef, J., \& Storchi-Bergmann, T.  2000, ApJ, 533, 682

\bibitem[]{504} Conselice, C.J. 2003, ApJS,147,1

\bibitem[]{506} Carroll, S.M., Press, W.H., \& Turner, E.L. 1992,
ARAA, 30, 499

\bibitem[]{819} Chabrier, G. 2003, PASP, 115, 763

\bibitem[]{511} Conselice, C., Bershady, M., Dickinson, M., \&
Papovich, C. 2003, AJ, 126, 1183

\bibitem[]{514} Coe, D., Benitez, N., Sanchez, F., Jee, M., Bouwens, R., \& Ford, H. et al. 2006, AJ, in press

\bibitem[]{516} Cowie, L., Hu, E., \& Songaila, A. 1995, AJ, 110,
1576

\bibitem[]{} de Mello, D., Wadadeker, Y., Dahlen, T., Casertano,
S., \& Gardner, J.P. 2006, AJ, 131, 216

\bibitem[]{} de Vaucouleurs, G., \& Buta, R. 1980, AJ, 85, 637

\bibitem[]{519} Elmegreen, B.G., \& Elmegreen, D.M. 2005, ApJ, 627,
632

\bibitem[]{522}Elmegreen, D.M., Elmegreen, B.G., \& Sheets, C.
2004,ApJ, 603, 74

\bibitem[]{525} Elmegreen, D.M., Elmegreen, B.G., Ravindranath, S.,
\& Coe, D. 2006, ApJ, in preparation

\bibitem[]{528} Elmegreen, D.M., Elmegreen, B.G., Rubin, D.S., \&
Schaffer, M.A. 2005a, ApJ, 631, 85

\bibitem[]{531} Elmegreen, B., Elmegreen, D., Vollbach, D., Foster,
E., \& Ferguson, T. 2005b, ApJ, 634, 101

\bibitem[]{} Few, J.M.A., \& Madore, B.F. 1986, MNRAS, 222, 673

\bibitem[]{534} Fosbury, R.A.E., \& Hawarden T.G. 1977, MNRAS, 178,
473

\bibitem[]{} Fukugita, M., Shimasaku, K., \& Ichikawa, T. 1995,
PASP, 107, 945

\bibitem[]{537} Giavalisco, M., et al. 2004, ApJ, 600, L103

\bibitem[]{539} Hernquist, L., \& Weil, M. 1993, MNRAS, 261, 804

\bibitem[]{541} Higdon, J., \& Wallin, J. 1997, ApJ, 474, 686

\bibitem[]{543} Huang, S.-N., \& Stewart, P. 1988, A\&A, 197, 14

\bibitem[]{545} Korchagin, V., Mayya, Y.D., \& Vorobyov, E. 2001,
ApJ, 554, 281

\bibitem[]{548} Lavery, R., Remijan, A., Charmandaris, V., Hayes,
R., \& Ring, A.  2004, ApJ, 612, 679

\bibitem[]{878} Leitherer, C., Li, I.-H., Calzetti, D., Heckman,
T.M. 2002, ApJS, 140, 303

\bibitem[]{554} Lotz, J.M., et al. 2006, astro-ph/0602088

\bibitem[]{556} Lubin, L. M., \& Sandage, A. 2001, AJ, 122, 1084

\bibitem[]{558} Lynds, R., \& Toomre, A. 1976, ApJ, 209, 382

\bibitem[]{881} Madau, P. 1995, ApJ, 441, 18

\bibitem[]{562} Mihos, C., \& Hernquist, L. 1994, ApJ, 437, 611

\bibitem[]{564} Neuschaefer, L., Im, M., Ratnatunga, U., Griffiths,
R., \& Casertano, S. 1997, ApJ, 480, 59

\bibitem[]{567} Rasband, W.S., 1997, ImageJ, U.S. National
Institutes of Health, Bethesda, MD, http://rsb.info.nih.gov/ij/

\bibitem[]{570} Rix, H.W., et al. 2004, ApJS, 152, 163

\bibitem[]{902} Rowan-Robinson, M. 2003, MNRAS, 345, 819

\bibitem[]{} Schweizer, F., Ford, W.K., Jedrzejewski, R., \&
Giovanelli, R. 1987, ApJ, 320, 454

\bibitem[]{574} Smith, B.J., Struck, C., Appleton, P.N.,
Charmandaris, V., Reach, W., \& Eittier, J.J. 2005, AJ, 130, 2117

\bibitem[]{577} Spergel, D. N., et al. 2003, ApJS, 148, 175

\bibitem[]{579} Straughn, A., Cohen, S., Ryan, R., Hathi, N.,
Windhorst, R., \& Jansen. R. 2006, ApJ, 639, 724

\bibitem[]{582} Struck, C. 1997, ApJS, 113, 269

\bibitem[]{584} Struck, C. 1999, Physics Reports, 321, 1

\bibitem[]{586} Struck. C., Appleton, P., Borne, K., \& Lucas, R.
1996, AJ, 112, 1868

\bibitem[]{589} Struck-Marcell, C. \& Appleton, P. 1987, ApJ, 323,
480

\bibitem[]{592} Struck-Marcell, C. \& Higdon, J.L. 1993, ApJ, 411,
108

\bibitem[]{595} Theys, J.C., \& Spiegel, E.A. 1976, ApJ, 208, 650

\bibitem[]{597} Theys, J.C., \& Spiegel, E.A. 1977, ApJ., 212, 616

\bibitem[]{599} Thompson, L., \& Theys, J. 1978, ApJ, 224, 796

\bibitem[]{} Whitmore, B. C., Lucas, R. A., McElroy, D. B.,
Steiman-Cameron, T. Y., Sackett, P. D., \& Olling, R. P. 1990, AJ,
100, 1489

\bibitem[]{} Windhorst, R.A., et al. 2002, ApJS, 143, 113

\bibitem[]{601} Wolf, C., Meisenheimer, K., Rix, H.-W., Borch, A.,
Dye, S., \& Kleinheinrich, M. 2003, A\&A, 401, 73

\bibitem[]{604} Wong, O., Meurer, G., Bekki, K., et al. 2006, MNRAS,
in press

\end{thebibliography}
\end{document}